\documentclass[overfull]{epl}
\usepackage{amsmath}
\usepackage{amsbsy}
\usepackage{amssymb}
\usepackage{psfrag}
\newcommand{\corresponds}{\equiv}
\newcommand{\arccosh}{\ab{arccosh}}
\newcommand{\openone}{\leavevmode\hbox{\small1\kern-2.8pt\normalsize1}}
\title{Wave chaos in elastodynamic cavity scattering}
\author{A. Wirzba\inst{1}
\and
N. S\o ndergaard\inst{2}
\and
P. Cvitanovi\'c\inst{3}} 
\shortauthor{A. Wirzba \etal}
\institute{
   \inst{1} Institut f\"ur Kernphysik,
               Forschungszentrum J{\"u}lich -
               D-52425 J{\"u}lich, Germany\\ 
  \inst{2} Matematisk Fysik, 
                Lunds Universitet -
                Box 118,
                SE-22100 Lund,
                Sweden\\
  \inst{3} Center for Nonlinear Science, School of Physics,
                Georgia Institute of Technology -
                Atlanta, GA 30332-0430, USA
}

\begin{document}

\maketitle

\begin{abstract}
  The exact elastodynamic scattering theory is 
  constructed to describe the \mbox{spectral} properties of
  two- and more- 
  cylindrical cavity systems,
  and compared to an elastodynamic
  generalization of the semi-classical Gutzwiller unstable periodic orbits
  formulas.
  In contrast to quantum mechanics, 
  %here 
  complex periodic orbits associated with 
  the surface Rayleigh waves dominate 
  the 
  low-frequency spectrum, 
  and already the two-cavity system displays 
  chaotic features.
  %The shortest of these orbits suffice to interpret 
  %the Wigner time delays and
  %the leading elastodynamic scattering resonances.
\end{abstract}

%----------------------------------------------------------
\section{Introduction}
%----------------------------------------------------------

The Gutzwiller semi-classical quantization of classically chaotic systems
relates quantum observables such as spectral densities to sums over classical
unstable periodic orbits~\cite{gutbook,DasBuch}.  The work presented here is a
step toward a formulation of such approximate short-wavelength  
theory of wave
chaos for the case of linear elastodynamics.  Why elastodynamics?  The
experiments 
initiated in ref.\,\cite{oxborrow}
attain
$Q$~values as high as $5 \cdot 10^6$, making spectral measurements in
elastodynamics competitive with measurements in microwave cavities at liquid
helium temperatures~\cite{heliumMicrowave1,heliumMicrowave2}, 
and vastly superior 
to nuclear
physics and room temperature microwave experiments for which the $Q$ values
are orders of magnitude lower, typically $\sim 10^{2}$--$10^{3}$.
For elastodynamics there are only 
a few experimental demonstrations~\cite{Kudrolli02}  of
the existence of unstable periodic orbits, and 
no theory that would predict them.
While Oxborrow \etal {\cite{oxborrow}}  
measure about $10^5$ spectral lines,
the current theory is barely adequate for computation of dozens of resonances.
A more effective 
theory would find many applications
such as in the frequency domain quality testing 
for small devices built from high $Q$ materials.
This unsatisfactory 
state of affairs is the  {\em raison d'\^{e}tre}  for the
theoretical effort undertaken here.

While current experiments excel in measurements of eigenspectra of 
{\em  compact resonators\/}, 
the periodic orbit theory computations of such bound
system spectra are rendered difficult by the 
presence of non-hyperbolic phase
space regions.  As our primary goal is to derive and test rules for replacing
wave mechanics by the 
short-wavelength ray-dynamic trajectories, we concentrate
here instead on the problem of {\em scattering} off cylindrical cavities, for
which the classical dynamics is fully under control.
In the case of one cavity the exact scattering spectrum is
known~\cite{Izbicki}. 
For the multiple-cavities case 
we generalize the
quantum-mechanical (QM) S-matrix formalism for $N$-disk
scattering~\cite{Berry81,gasp,AW_creep,AW_report}, 
and compute the exact
resonances and the Wigner time delays from the full elastodynamic
wave-mechanical scattering matrix.  
We then compare the exact results with the
corresponding quantities calculated in the short-wavelength
approximation (SWA), and
discover that the QM intuition fails us: the Rayleigh surface
waves (which have no analog in the QM scattering problem) dominate the
low-frequency spectrum 
because of their weights and  number, 
such that 
already the two-disk elastodynamic scattering problem displays 
chaotic features in
this regime in contrast to its QM 
counter part.

%----------------------------------------------------------
\section{Elastodynamics}
%----------------------------------------------------------

Consider an infinite slab of an isotropic and homogeneous elastic material
({\it e.g.\/}, polyethylene or isotropic quartz)
with parallel top and bottom plane
boundaries, and an in-phase stimulus such that the system behaves
quasi--two-dimensionally along the slab, with no excitation of or coupling to
waves propagating perpendicular to the slab.
The propagating waves are 
either the pressure  or the shear 
solutions of the Navier-Cauchy equation~\cite{lAndl}
$
 \mu \boldsymbol\nabla^2 \vect{u}
 + (\lambda + \mu) {\boldsymbol \nabla} 
 \bigl( {\boldsymbol\nabla} \cdot\vect{u}\bigr)
 + \rho \omega^2 \vect{u} = \vect{0}
$, 
where
$\vect{u}$ is a vectorial displacement field, $\lambda$ and
$\mu$ are the Lam\'e constants, $\rho$ the mass density and $\omega$ the
frequency.  The experiments dictate free boundary conditions, with vanishing
%(vectorial)
traction $\vect{t}(\vect{u}) = \vect{0}$ where 
%----------------------------------------------------------
\begin{eqnarray}
\vect{t}(\vect{u}) \equiv 
 \biggl[\lambda \Bigl({\boldsymbol \nabla} \cdot \vect{u}\Bigr)  
\openone 
 + \mu \Bigl\{ \bigl({\boldsymbol \nabla} \vect{u}\bigr) 
 + \bigl({\boldsymbol \nabla} 
 \vect{u}\bigr)^{\mathsf{T}}\Bigr\}\biggr]
 \cdot \hat{\vect{n}} 
\,.
 \label{traction}
\end{eqnarray}
%----------------------------------------------------------
Here $\hat{\vect{n}}$ is a unit vector normal to the boundary, $\openone$ the
unit matrix, and $\mathsf{T}$ indicates a transposition.  
Elastodynamic waves are vectorial, with the pressure 
(longitudinal -- $\ab{L}$) wave
propagating through the bulk with velocity 
$ c_{\ab{L}} = \sqrt{({\lambda + 2 \mu)}{/\rho}}$, 
and the shear (transverse -- $\ab{T}$)  wave with velocity $c_{\ab{T}} =
\sqrt{{\mu}/{\rho}}$.  Furthermore, the Rayleigh surface waves propagate along
plane boundaries unattenuated, with the velocity $c_{\ab{R}}$
determined~\cite{lAndl} 
by the condition
%----------------------------------------------------------
\begin{equation}
\label{RayleighVeloc}
  0= \left (2 \eta^2 -1\right)^2
 -4 \eta^2 
 \sqrt{(\eta^2-1)(\eta^2-\left({c_{\ab{T}}}/{c_{\ab{L}}}\right)^2)}
 \qquad \tx{with}\quad \eta \equiv c_{\ab{T}}/c_{\ab{R}}\,.
\end{equation}
%----------------------------------------------------------
When either a shear or pressure plane wave hits a boundary, mode conversion
can take place~\cite{couch}, with waves of different types 
emitted at different
angles.
We now drill one, two, or more cylindrical cavities
of radius $a$
perpendicularly through
the slab.

%----------------------------------------------------------
\section{One-cavity scattering, the exact spectrum}
%----------------------------------------------------------

Scattering off a single cylindrical cavity is {separable} in angular
momentum. 
The one-cavity S-matrix of elastodynamics, a [$2 \times 2$]
matrix in $\{\ab{L},\ab{T}\}$ components, is determined by the free boundary
condition for the traction matrices
%----------------------------------------------------------
\begin{equation}
\label{S-1-cav}
 \tens{S}^{(1)}_{m m'} = 
 - \delta_{m m'}\, \bigl[\tens{t}^{(+)}_{m} \bigr]^{-1}
 \cdot
  \bigl[\tens{t}^{{(-)}}_{m}\bigr] 
\,,\qquad
 m = 0,\pm 1, \pm 2, \cdots\,. 
\end{equation}
%----------------------------------------------------------
For comparison, the corresponding one-disk scattering
matrix~\cite{franz,AW_report} of quantum mechanics (QM), for a 
scalar field with the Dirichlet boundary condition, is 
expressed in terms of Hankel 
functions,
${S}^{(1)}_{m m'} = 
 -\delta_{m m'} { \ab{H}_{m'}^{(-)}(ka)}/ {
 \ab{H}_m^{(+)}(ka)}$, as function of the wave number $k$.
The traction {\em matrices} \cite{NSthesis00} 
of angular momentum $m$ and superscript 
$Z\in\{+,-,\ab{J}\}$
result
when  outgoing $(+)$, incoming $(-)$ or regular $(\ab{J})$ 
pressure and shear displacements $\vect{u}$  in terms of
Hankel  or 
Bessel functions, $Z_m\in\{\ab{H}_m^{(+)}, \ab{H}_m^{(-)}, \ab{J}_m\}$, are
inserted into the traction (\ref{traction}): 
%----------------------------------------------------------
\begin{equation}
\label{t-def}
 \bigl[{\tens{t}^{(Z)}_m}\bigr]_{\pi  \sigma} =\frac{2\mu}{a^2}
 \left[(-1)^{\sigma}\delta_{\pi \sigma}  
 \left(k_\sigma \frac{\upd\ }{\upd k_\sigma}
 -\bigl(m^2- {\textstyle\frac{1}{2}}k_{\ab{T}}^2 a^2\bigr)\right)
-im (1-\delta_{\pi  \sigma })\, 
 \left(k_\sigma\frac{\upd\ }{\upd k_\sigma}-1\right) 
\right] 
 Z_m(ak_\sigma).
\end{equation}
%----------------------------------------------------------
Index  $\pi\in\{1,2\}$ labels the 2-$d$ 
spherical components 
$\{\hat{\vect{r}},{\hat {\boldsymbol \theta}}\}$ 
of the displacement $\vect{u}$ 
at the cavity
while 
$\sigma\in\{1,2\} $
labels its polarization  $\{\ab{L},\ab{T}\}$ 
where $k_\sigma=\omega/c_\sigma$ are
the corresponding wave numbers. 
The scattering resonances for the one-cavity elastodynamic 
medium with traction-free 
boundary conditions are determined~\cite{Izbicki} 
from
$ 
 {\ab{Det}} \bigl[\tens{t}^{(+)}_m  \bigr]=0
$, see eq.~(\ref{S-1-cav}).

%----------------------------------------------------------
\section{Multi-cavity scattering, exact spectrum}
%----------------------------------------------------------

We construct the
S-matrix for the $N$-cavity system following ref.~\cite{AW_report}.
The determinant of the $N$-scatterer S-matrix 
factorizes,
as in QM, into the product of the one-cavity determinants
and the multi-scattering contributions~\cite{AW_report,NSthesis00}:
%----------------------------------------------------------
\begin{equation}
 \label{det-S-N-cav}
 \det{} \tens{S}(\omega)
 =       
  \frac{ {\ab{Det}}\bigl[\tens{M}(\omega^\ast)^\dagger \bigr ]}
 { {\ab{Det}}\bigl[\tens{M}(\omega) \bigr]}\,
 \prod_{{j}=1}^{N}\det{} {\tens{S}^{(1)}}^j(\omega)
\,.
\end{equation} 
%----------------------------------------------------------
Here the matrix $\tens{M}= \openone+ \tens{A}$ 
is the inverse of the multi-scattering matrix.
The  transfer matrix $\tens{A}$ evolves the
displacement 
of internal angular momentum  $l'=0,\pm 1,\pm 2, \cdots$ 
and spherical component
$\pi'\in\{1,2\}$ at cavity~$j'$ (of radius $a_{j'}$) to the 
($l=0,\pm 1,\pm 2, \cdots$ and $\pi\in\{1,2\}$) displacement at
cavity~$j$ (of radius $a_j$) where $j,j'=1,\cdots,N$ are the cavity-labels:
%----------------------------------------------------------
\begin{eqnarray}
 \bigl[\tens{A}^{jj'}_{ll'}\bigr]_{\pi\pi'} 
  &=&(1-\delta_{jj'})
  \frac{a_j}{a_{j'}}\sum_{\sigma=1}^2 \sum_{\sigma'=1}^2
\bigl[ {\tens{t}_l^{(J)j}} \bigr]_{\pi\sigma}
  \Bigl[{\tens{T}_{ll'}^{(+)jj'}}
\Bigr]_{\sigma\sigma'} 
  \bigl[ {\tens{t}_{l'}^{(+)j'}} \bigr]^{-1}_{\sigma'\pi'}\,.
\end{eqnarray}
%----------------------------------------------------------
The matrix 
$
 [ {\tens{T}_{ll'}^{(+) jj' }}]_{\sigma \sigma'}$= 
  $\delta_{\sigma \sigma'} 
  \ab{H}_{l-l'}^{(+)}(k_{\sigma} R_{jj'}) \exp\left[il \alpha_{j'j}- 
  il'(\alpha_{j j'}-\pi)\right]
$ 
translates the $\{\ab{L},\ab{T}\}$ modes evaluated relative 
to the origin of 
cavity $j'$ to the corresponding modes at cavity $j$.
$R_{jj'}$
 and $\alpha_{jj'}$ are the relative center-to-center distances and
angles, respectively~\cite{AW_report,NSthesis00}.
The multi-scattering resonances of the
$N$-cavity problem
are given by the zeros of ${\ab{Det}}\, \tens {M}(\omega)$, see 
eq.~(\ref{det-S-N-cav}).
For the two-cavity system the two-fold reflection
symmetry implies that the determinant factors into four
irreducible representations,
with  the transfer matrix for each irreducible subspace
defined on the fundamental domain, a quarter 
of the full elastodynamic slab~\cite{DasBuch}.

\begin{figure} % of fig.1
 \psfrag{Im k_L a}{\large Im $k_{\ab{L}} a$}
 \psfrag{Re k_L a}{\large Re $k_{\ab{L}} a$}
 \psfrag{i-lead}{\em i-lead}
 \psfrag{ii-lead}{\em ii-lead}
 \psfrag{i-add}{\em i-add}
 \psfrag{ii-sub}{\em ii-sub}
 \psfrag{iii-sub}{\em iii-sub}
 \psfrag{ii-ris}{\em ii-ris}
 \psfrag{iii-ris}{\em iii-ris}
 \psfrag{iv-ris}{\em iv-ris}
 \psfrag{iii-ris}{\em iii-ris}
 \psfrag{iii-reg}{\em iii-reg}
 \psfrag{iv-reg}{\em iv-reg}
 \psfrag{iii-drop}{\em iii-drop}
 \onefigure[width=6.5cm,angle=270]{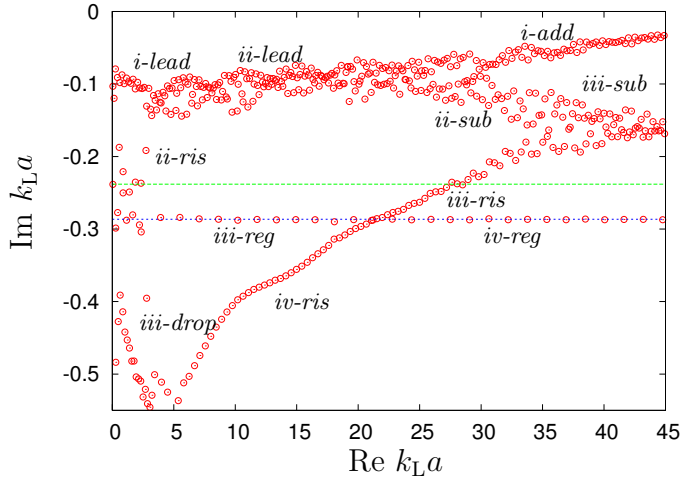}
 \caption{
The lowest few hundred exact $A_1$ resonances of the two-cavity
elastodynamic scatterer, in
the complex pressure wave number $k_{\ab{L}}=\omega/ c_{\ab{L}}$ plane.
The labels refer to the various resonance families interpreted
(together with the lines) in
the {\em Discussion} section, where the  italics specify 
the orders in the cumulant expansion at which the structures are first 
observed. The second label
indicates whether the bands are {\em lead\/}ing (closest to the real axis), 
{\em add\/}itional,
{\em sub\/}leading, {\em ris\/}ing, {\em reg\/}ular or {\em drop\/}ping. 
        }
\label{fig:A1_res}
\end{figure}

Here we present typical numerical results for 
the fully symmetric $A_1$ subspace of
the system of two cylindrical cavities.
In all presented calculations  we
take values of the Lam\'e constants corresponding 
to polyethylene~\cite{Izbicki},
and set $c_{\ab{L}}$ = $1950\un{m/s}$,
$c_{\ab{T}}$ = $540\un{m/s}$, $c_{\ab{R}}\approx 513\un{m/s}$,
and take cavities of radius $a$ = $1\un{cm}$, 
center-to-center separation  $R$ = $6\un{cm}$.
The  lowest few hundred 
exact $A_1$ resonances 
(hopefully, {\em all} $A_1$ resonances
       in the window
      [$0 < \ab{Re}\, k_{\ab{L}} a < 45$,$-0.55 <\ab{Im}\, k_{\ab{L}} a < 0$]) 
determined by the zeros of 
$\left.{\ab{Det}}(\openone+\tens{A})\right|_{A_1}$
are shown in fig.~\ref{fig:A1_res}.

%----------------------------------------------------------
\section{One-cavity scattering, ray-dynamic interpretation}
%----------------------------------------------------------

In the one-cavity case a sophisticated theory of ray dynamics already exists.
The main tool is the Sommerfeld-Watson transformation which in the QM case
replaces a slowly converging partial wave sum by a fast converging sum over
{\em complex creeping} trajectories, first derived by Franz~\cite{franz}.
The pressure and shear Franz resonances
also exist in elastodynamics~\cite{kellerRayleigh}, but here
the one-cavity spectrum is dominated
by the weakly damped Rayleigh resonances\,\cite{Izbicki}, 
with no QM counterpart.

In the spirit of Keller's geometrical theory of
diffraction~\cite{kellerRayleigh} we use the
one-cavity S-matrix (\ref{S-1-cav})
to assign a ray-dynamic weight to a segment of the boundary traversed by a
Rayleigh wave.  A circular Rayleigh segment of arc length $\Delta \phi\, a$ 
(with $\Delta\phi$ the pertinent angle) has
the complex-valued weight
$\exp[i \Delta\phi\, \nu_{\ab{R}}(\omega)]$, such that the effective
arc length of a Rayleigh segment is 
$
 \Delta\phi\,\nu_{\ab{R}}(\omega)/k_{\ab{T}} \approx \Delta \phi  \,a \eta
$,
where $\eta = c_{\ab{T}}/c_{\ab{R}}$, see eq.~(\ref{RayleighVeloc}).
Here we take $\Delta\phi$ times the {\em exact wave-mechanical} 
$\nu_{\ab{R}}(\omega)$,  determined as complex angular momentum by 
${\ab{Det}}[
\tens{t}^{(+)}_{\nu_{\ab{R}}(\omega)} ]=0$,
as the {\em input} for the complex-valued action of the
Rayleigh trajectory segment.

%----------------------------------------------------------
  % of fig. 1
%----------------------------------------------------------

%----------------------------------------------------------
\section{Multi-cavity scattering, ray-dynamic interpretation} 
%----------------------------------------------------------

In QM
the Gutzwiller-Voros Zeta function is
the semi-classical approximation to ${\ab{Det}}\, 
\tens{M}$, where the
connection follows via the semi-classical reduction of the traces
${\ab{Tr}}(\tens{A}^n)$ 
appearing in the cumulant expansion 
$
%\begin{equation}
 {\ab{Det} }\,\tens{M} 
  = 1 +{\ab{Tr}}\,\tens{A}
  -\left[ {\ab{Tr}}\,\tens{A}^2 - ({\ab{Tr}}\,\tens{A})^2
      \right]/2 + \cdots 
%\end{equation}
$.
As shown in refs.~\cite{AW_creep,AW_report}, 
in the SWA 
(applicable here  from $k_{\ab{L}}a$ $\sim$ 2 onwards)
the traces ${\ab{Tr}} \tens{A}^n$ reduce to the set of
ray-dynamic periodic orbits of topological length $n$, 
whereas the cumulants become  ``curvatures'' (periodic orbits shadowed by
pseudo orbits~\cite{DasBuch}). 
The shortest geometrical periodic orbits, 
bouncing between the two cavities, are the border orbits 
of topological length one
in the fundamental domain.
Their weights, derived from the SWA 
to ${\ab{Tr}}\,\tens{A}_{A_1}$,
are of form\,\cite{AW_report,NSthesis00}
$
 t_{\sigma} =  -
\exp\bigl[i k_\sigma (R-2a)\bigr]/\bigl[\sqrt{|\Lambda_0|}
(-\Lambda_0)^{\sigma-1} 
   (1 -\Lambda_0^{-2})\bigr]$
 with 
$\Lambda_0 = \bigl(R-a +\sqrt{R^2-2Ra}\bigr)/a$,
$\sigma \in\{1,2\}\corresponds\{\ab{L},\ab{T}\}$.
These unstable geometrical orbits including 
their repetitions are summed up in the usual way in terms of the
Gutzwiller-Voros spectral determinant 
${\ab{Det}}\,\tens{M}_{\ab{geom.}}$\,\cite{DasBuch}.

\begin{figure} % of fig. 1
% shrunk this \onefigure[height=2.7cm]{rayleigh_two.eps}
 \onefigure[height=2.1cm]{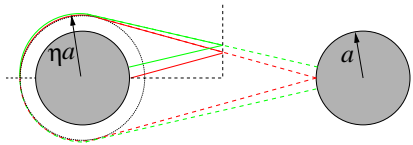}
 \caption{
Two new periodic orbits of Rayleigh type of topological length 2 
(and ``effective'' arc radius $\eta a > a$)
in the fundamental domain (and the full space)
of the two-cavity system. 
In the former 
the Rayleigh orbits are further classifiable 
by the number of  contacts with the two 
symmetry  axes~\cite{VattWirzRos,RosVattWirz}.
 }
\label{fig:Rayl_orbits}
\end{figure}

Moreover,
Keller's theory
yields
orbits with unstable geometrical legs {\em and} 
weakly damped Rayleigh surface wave arcs circling the cavities.
At the topological length one,
two (unstable) complex Rayleigh-type periodic orbits, 
an ``oval'' and ``figure eight''
contribute~\cite{AW_creep}.
In addition to the repeated primary 
orbits {\em and} the ``oval--eight'' combination, 
two new (unstable) Rayleigh-type orbits contribute to
${\ab{Tr}}\,\tens{A}^2$, see fig.~\ref{fig:Rayl_orbits}
and so on for longer Rayleigh-type orbits, 
with their total number (including repeats) 
growing exponentially with their
topological length.
%----------------------------------------------------------
 % of fig. 2
%----------------------------------------------------------
By analogy to creeping orbits~\cite{VattWirzRos,RosVattWirz},
if $q_i$, $i=1,\cdots,n$
(with $q_{n+i}\equiv q_i$) are the
points along
a cycle where a Rayleigh arc segment connects to 
geometric trajectories 
(which may include reflections with/without conversions between
pressure and shear rays),
the Rayleigh surface
wave contribution to the spectral determinant 
is of form
%----------------------------------------------------------
\begin{equation}
        \left.{\ab{Det} }\,\tens{M}\right|_{\ab{Rayl.}}
        = 
   \exp    \sum_{p,r=1}^\infty
\left({-1}/{r}\right)
        \left( \prod_{i=1}^{n_p} [G(q_{i+1},q_i)]_{\sigma_{i+1}\sigma_i}
    \right)^r 
\,. 
\end{equation} 
%----------------------------------------------------------
Here the  summation goes over all prime periodic orbits
$p$ and their repetition number $r$.
$[G(q_{i+1},q_i)]_{\sigma_{i+1}\sigma_i}$ 
(with $\sigma_{i},\sigma_{i+1}\in\{\ab{L},\ab{T}\}$)  is
the Van-Vleck propagator  (including instabilities, reflections 
and mode conversions\,\cite{NSthesis00,disc})
if $q_i$ and $q_{i+1}$ are connected by a pure 
geometric trajectory, and 
$\frac{i}{2} [\widetilde{\tens{D}}_{\ab{R}}(\omega)]_{\sigma_{i+1} \sigma_i}
              \exp(i\Delta\phi\,\nu_{\ab{R}}(\omega))$ if
$q_i$ and $q_{i+1}$ are the endpoints of a Rayleigh arc segment,
with $\sigma_i$, $\sigma_{i+1}$ the polarizations of the two attached
geometrical legs.
The [$2\times 2$]
diffraction matrix $\widetilde{\tens{D}}_{\ab{R}}(\omega)$ 
is the elastodynamics
analogue of the {\em square} of the corresponding QM
diffraction constant~\cite{VattWirzRos,RosVattWirz,AW_report}. 
It is
determined  by Keller and 
Karal~\cite{kellerRayleigh} for the half-plane case, and
in \cite{NSthesis00} for the circular cavity. 
$[\widetilde{\tens{D}}_{\ab{R}}(\omega)]_{\sigma_{i+1} \sigma_i}$ is  
proportional to  
$\exp[-\lambda(\omega)_{\sigma_{i+1}}
-\lambda(\omega)_{\sigma_i}]$ with
$\lambda(\omega)_{\sigma}\equiv 
\nu_{\ab{R}}\arccosh[\nu_{\ab{R}}/(a k_\sigma)]
-\sqrt{\nu_{\ab{R}}^2-(a k_\sigma)^2}$,
{\it i.e.\/}, it is weakly attenuated for the pure shear case 
$\sigma_{i+1}$ = $\sigma_{i}$ = $\ab{T}$ at low frequencies, but 
strongly attenuated otherwise.

The ``semiclassical'' form of the spectral determinant 
is given by the formal product
${\ab{Det}}\,\tens{M}|_{\ab{geom.}} \times 
{\ab{Det}}\,\tens{M}|_{\ab{Rayl.}}$,
evaluated in the cycle expansion where the  
classifying topological length is
equal to the number of geometrical straight legs of the orbits  and
pseudo orbits\,\cite{DasBuch}.

\begin{figure} % of fig. 3
 \psfrag{k}{\!\large$k$}
 \psfrag{L}{$\ab{L}$}
 \psfrag{ a}{\large$a$}
 \onefigure[width=7.4cm]{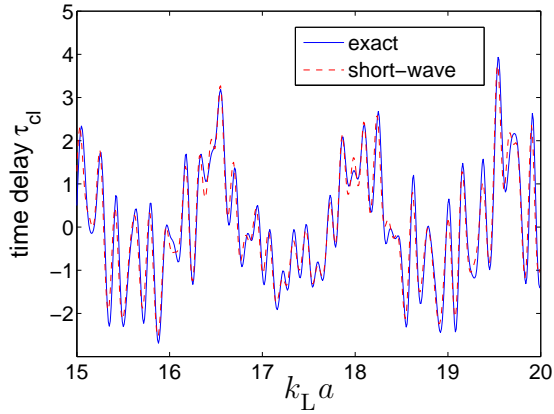}
 \caption{
  Wigner time delay for the two-cavity system,
  the exact $A_1$ result  
  versus its ray-dynamic expansion to period two,
  as function of the pressure wave number $k_{\ab{L}}$ (times the
cavity radius $a$).
        }
 \label{fig:delay_vs_k}
\end{figure}

\begin{figure} % of fig.4 left/right
 \twoimages[width=7.1cm]{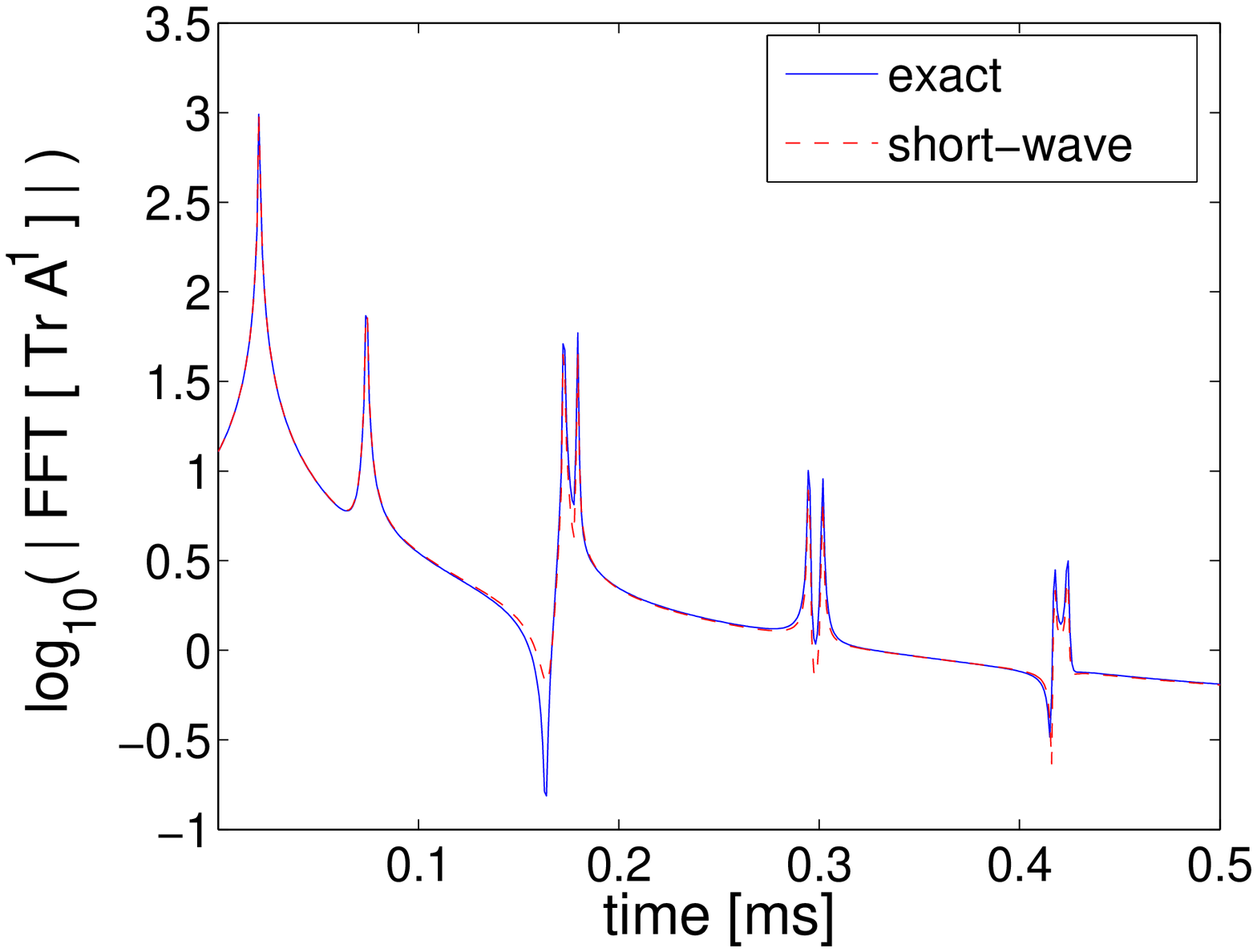}{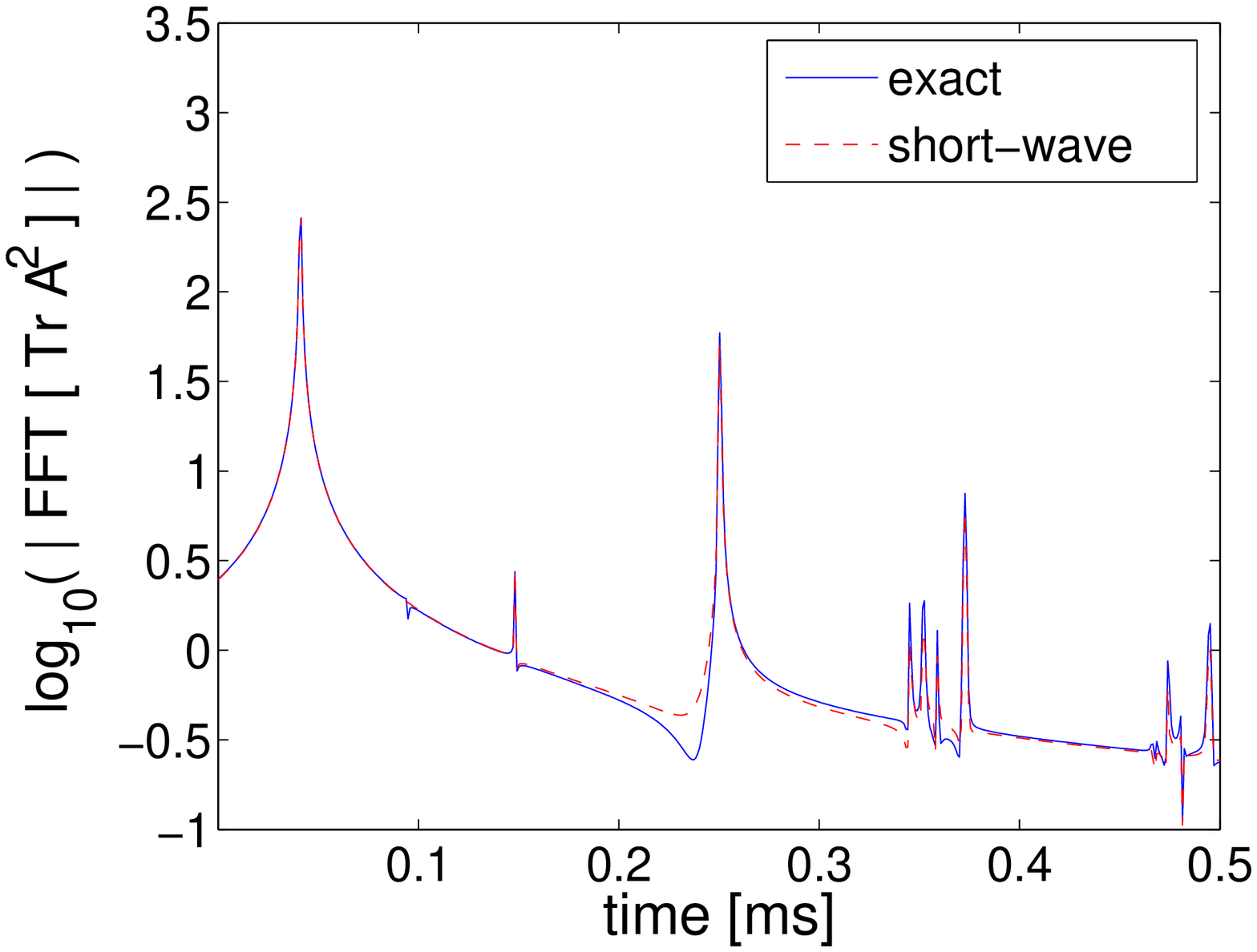}
 \caption{
 Comparison of the Fourier transforms 
 of the exact traces
 ${\ab{Tr}}\, \tens{A}_{A_1}$ (left panel) and
 ${\ab{Tr}}\, \tens{A}^2_{A_1}$ (right panel)
 of the two-cavity system  
 with the short-wavelength approximation (SWA) based on 
 periodic orbits of topological length one and 
 periodic orbits of topological length two, respectively, 
 in the range $10\leq k_{\ab{L}} a\leq 45$. All peaks are
 identifiable as the periods of the shortest period orbits.}
 \label{fig:FFT_tra}
\end{figure}

The  Wigner time delays and cumulant traces are 
particularly well suited to detailed comparisons
of the exact results with the SWA~\cite{AW_report}.
In fig.~\ref{fig:delay_vs_k} we plot the exact Wigner ``time'' delay
$\tau_{\ab{cl}} = \frac{\upd\ \ }{\upd k_{\ab{L}} a} 
\eta_{\ab{cl}}$ 
of the cluster phase shift
$
\eta_{\ab{cl}}(\omega) 
= -i{\textstyle\frac{1}{2}}\ln\bigl[
  {\ab{Det}}\, \tens{M}(\omega^\ast)^\dagger/{\ab{Det}}\,
\tens{M}(\omega)\bigr]
$
as function of the pressure wave number $k_{\ab{L}}=\omega/c_{\ab{L}}$, 
and compare it
to the cycle 
expansion based on the 
geometrical and
complex periodic orbits of topological lengths one and two.
The periodic orbit approximation is in
good agreement with the exact 
result which can be truncated
at second cumulant order for the presented $k_{\ab{L}}$ values.
%----------------------------------------------------------
 % of fig.3
%----------------------------------------------------------
Finally,
in fig.~\ref{fig:FFT_tra}
we compare the 
periods of the ray-dynamic periodic orbits 
with the Fourier peaks of the exact wave-mechanical data
by plotting the
moduli of the Fourier transforms of ${\ab{Tr}}
\tens{A}_{A_1}$
and ${\ab{Tr}}\,\tens{A}^2_{A_1}$ for the region $10\leq k_{\ab{L}} a \leq 45$,
and 
the corresponding Fourier transforms of sums over
periodic orbits 
of topological length one and two, respectively.
The 
Fourier peaks indeed correspond to
the periodic-orbit periods
$
T_{{p}} \,=\, \sum_{i_{\ab{L}}} l_{i_{\ab{L}}} /c_{\ab{L}} 
 + \sum_{i_{\ab{T}}} l_{i_{\ab{T}}} /c_{\ab{T}}
        + \sum_{i_{\ab{R}}} \eta a\Delta\phi_{i_{\ab{R}}}/c_{\ab{T}}
$,
with $l_{i_{\ab{L}}}$, $l_{i_{\ab{T}}}$ and $\eta a \Delta\phi_{i_{\ab{R}}}$ 
the (effective) geometrical lengths
of the pressure, shear and Rayleigh segments, respectively.

The SWA captures nearly all
qualitative features of the exact calculation. 
However, the $1/k_{\ab{L}}a$ corrections are still not negligible, as
can be seen from the small structure close to $0.1\un{ms}$ in 
fig.~\ref{fig:FFT_tra} (right).
This corresponds to a combined $\ab{L} \ab{T}$ geometric orbit  that
vanishes in the SWA, as there is no coupling
of pressure to shear rays at perpendicular impact.
%----------------------------------------------------------
 % of fig.4 left/right
%----------------------------------------------------------
The 
complex orbits
visible in fig.~\ref{fig:FFT_tra}
have only shear segments coupled to
Rayleigh segments,
whereas the coupling
of pressure segments to Rayleigh segments is
severely suppressed (since $c_{\ab{L}} \gg c_{\ab{R}}$).

%----------------------------------------------------------
\section{Discussion}
%----------------------------------------------------------

These observations combined with the cumulant expansion
(in detail beyond the scope of this letter)
explain the qualitative features of the low-frequency resonance spectrum
shown in  fig.~\ref{fig:A1_res}.
The dotted and dashed lines represent the imaginary parts
$
\ab{Im}\, k_{\ab{L}}$  = $-{\textstyle\frac{1}{2}}\ln (\Lambda_0) 
/ (R-2a)$ $\approx -0.29/a$ 
and 
$
\ab{Im}\, k_{\ab{T}}  = -{\textstyle\frac{3}{2}}\ln (\Lambda_0) 
c_{\ab{T}} / c_{\ab{L}}
 (R-2a)\approx -0.24/a
$ 
predicted by the  isolated 
pressure $t_{\ab{L}}$ and
shear $t_{\ab{T}}$ orbit, respectively.
As the transverse oscillation of the $t_{\ab{T}}$
orbit strongly mixes with the Rayleigh waves of nearly the same
wave velocity,
the  irregular  band with the {\em leading}  
and
{\em subleading} resonances
arises from the interference of the
Rayleigh orbits with the 
$t_{\ab{T}}$ orbit.
Due to this mixing,
there are no resonances on the $t_{\ab{T}}$ line.

The Rayleigh orbits include the factor
$( 1 - \exp ( 2 \pi i \nu_{\ab{R}}) )^{-1}$ which arises from
the geometrical sum of the {\em additional} Rayleigh
waves around the half-cavity in the fundamental domain. These 
1-cavity structures, which lead to the leading 
resonances in the 1-cavity case, 
correspond here 
to poles in  $\det M(k)$ which amplify the contributions of the Rayleigh
orbits already
at topological length one, but which also 
complicate the searches for nearby genuine multi-scattering resonances,
the zeros of $\det M(k)$. This is especially the case for the
family {\em i-add}. 

Below  ${\ab{Re}} k_{\ab{L}} a \approx 32$ SWA cycles 
of at most topological length two are needed to get a qualitative fit
of the chaotic band labelled as {\em ii-lead}, 
whereas already order one is sufficient  
for the resonances {\em i-lead}, before they merge with the rising
band {\em ii-ris} at about ${\ab{Re}} k_{\ab{L}} a \approx 4$, and for the 
resonances {\em i-add}. Below  ${\ab{Re}} k_{\ab{L}} a \approx 2$, the SWA
expansion breaks down and a uniform approximation based on a multipole
expansion as in Ref.~\cite{RWW96} should take over.
The dropping  band, {\em iii-drop}, has a QM analog 
which is
generated  by Franz' creeping waves.
The {\em regular} family of resonances 
(see {\em iii-reg} and {\em iv-reg})
corresponds to the 
$t_{\ab{L}}$
orbit 
that dominates the QM two-disk case \cite{VattWirzRos}, 
but is subdominant here, and that couples neither 
to the complex Rayleigh orbits nor to the shear orbit.
When the {\em rising band} ({\em iv-ris}) crosses the regular band 
({\em iii-reg}), the cumulant orders three and four are interchanged.
It continues as band {\em iii-ris} and
merges with the subleading band {\em ii-sub}
at about ${\ab{Re}} k_{\ab{L}} a =35$, generating the combined band 
{\em iii-sub}. 

For the $k_{\ab{L}}$ window shown in fig.\,\ref{fig:A1_res}, the
periodic orbit sum can be truncated at length four,
since the spectrum generated
from the cumulant expansion to this order does not differ from
the complete one to the resolution of this figure. 
This is confirmed by 
numerical diagonalizations of the 
$M\times M$  matrices $\tens{M}_{A_1}(k_{\ab{L}})$ 
in the
$\{\ab{L},\ab{T}\}\,\times$
angular momentum 
space where   $M$
has to satisfy the bound $M > e ( c_{\ab{L}}/  c_{\ab{T}}) 
|k_{\ab{L}}| a$  ({\it e.g.\/}, 
$M > 440$ for $\ab{Re}\,k_{\ab{L}}  =45/a$) \cite{Berry81,AW_report}.

%----------------------------------------------------------
\section{Summary and outlook}
%----------------------------------------------------------

We have derived the scattering determinant and
calculated, in the low-frequency regime, 
the  exact scattering resonances and 
Wigner time delays  for a quasi-two-dimensional isotropic and
homogeneous elastodynamic slab with two cylindrical cavities. 
Already 
the physics of this simplest possible multi-scattering system in
elastodynamics with free
boundary conditions
is totally different
from the one of quantum billiards at low frequencies:
none of 
the {\em measurable} medium excitation ({\it e.g.\/}, phase shifts, 
leading resonances) can be understood without the Rayleigh waves which  do not
have quantum mechanical counterparts.             
For circular  boundaries they
are barely damped at all \cite{Izbicki} and the diffraction constants, which
link  them at the different cavities  to the shear waves,
are still only weakly attenuated  at low frequencies. These features  are
generic 
for smooth finite-size concave cavities.
The pressure waves which are {\em the} analog of the scalar 
quantum mechanical ones
play only a secondary role.

A symbolic dynamics  in the total space needs  to
account
for cycles patched together from 4 kinds of segments: 
pressure, shear (their sum  is the topological length),
anti- and clockwise Rayleigh,
implying an 
exponentially growing number of interfering
periodic orbits. % and thus a chaotic repeller. 
Whether the topological  increase  or the attenuation of the  Rayleigh 
orbits 
eventually 
wins is still open. 
Surface  orbits 
of Rayleigh type - with no counterpart in  QM -  
are expected in general  non-convex  elastic resonators.
Generalizations to anisotropic media 
(the highest $Q$-value experiments are performed on single
crystals of quartz), and applications of the above ray-dynamics
techniques to resonator geometries used in experiments 
remain open problems.
       
%----------------------------------------------------------

\end{document}